# Hunting for Room Temperature Superconductors

Huiqian Luo[1][*]

[1]Beijing National Laboratory for Condensed Matter Physics, Institute of Physics, Chinese Academy of Sciences, Beijing 100190, China
[*] Correspondence: hqluo@iphy.ac.cn

Superconductivity is the first discovered macroscopic quantum phenomenon since 1911. Superconductors are defined as the materials host both zero resistance and full diamagnetism (Meissner effect) states below the critical temperature ($T_c$) of a second order thermodynamic phase transition. In microscopic view, superconducting state is a condensation of coherent electron pairs (Cooper pairs) protected by finite energy gaps near the Fermi levels. Therefore, superconductors can be used as the hosts of strong electric current and high magnetic field, as well as quantum devices and chips, to break the limits of electromagnetic and quantum applications in traditional materials [1].

To date, thousands of superconductors have been discovered in many chemical forms such as pure elements, alloys, intermetallic compounds, heavy fermions, hydrides, cuprates, pnictides, nickelates, carbides even organic materials, etc (Fig.1). However, the superconducting applications are limited within several materials such as Nb, Nb-Ti, $Nb_3Sn$, $Nb_3Al$, $MgB_2$, $ReBa_2Cu_3O_{7-\delta}$, $Bi_2Sr_2CaCu_2O_{8+\delta}$, $Bi_2Sr_2Ca_2Cu_3O_{10+\delta}$, $Ba_{1-x}K_xFe_2As_2$, etc. None of them can fully replace the traditional cables or magnets due to their expensive costs. To search a useful superconductor, one must have high critical temperature, high upper critical field ($H_{c2}$) and high critical current density ($J_c$), nevertheless, it is better to show chemical stability, non-toxicity, flexibility, enough mechanical strength, low price, and easy production. Taking the extensively used Nb-Ti wire as an example, due to the limitation of $T_c$ < 10 K and $H_{c2}$ < 10 T, the Nb-Ti based magnets can be only used by cooling in liquid helium and under a medium field. Although the highest record of $T_c$ under ambient pressure is realized in $HgBa_2Ca_2Cu_3O_{8+\delta}$ system, this compound is toxic, sensitive in air and hard to synthesis. The iron-based superconductors open a new avenue for high field applications below 55 K, if its $J_c$ can be further improved in comparison with cuprates. So far, the REBCO ($ReBa_2Cu_3O_{7-\delta}$) tapes and Bi2212 ($Bi_2Sr_2CaCu_2O_{8+\delta}$) wires are the most promising superconducting materials for the applications under liquid nitrogen temperature (77 K), but the complex production technology results in high prices due to their intrinsic disadvantages on mechanical strength, extreme anisotropy of $H_{c2}$, weak pinning of magnetic flux and rich flux dynamics in the mixed state[2].

For decades, scientists desire to hunt for room temperature superconductors (RTS), whose $T_c$ are above the room temperature $T$=300 K. RTS retain the last hope of massive superconducting applications under ambient conditions, even though the $J_c$ and $H_{c2}$ may be not as expected. In fact, there are dozens of RTS reported in the last 50 years, but none of these results can be reproduced by other groups, thus they are known as "Unidentified Superconducting Objects" (USO). We must emphasis that the judgement of a superconductor is very strict on absolute zero resistance and near 100% diamagnetism below $T_c$ and under a low field. More importantly, the superconducting transitions is driven by the condensation of Cooper pairs, which means the entropy should be reduced accompanying by a jump of electron heat capacity. While all three criteria may not establish simultaneously, the drop down of resistivity, suppression of transitions under magnetic field, weak diamagnetism (< 1% volume) and anomaly in heat capacity cannot be alone taken as the evidence of superconductivity. Indeed, there are so many USO or RTS exhibiting spurious signatures such as a resistance transition induced by density waves, a diamagnetic signal due to contaminations, an anomaly in heat capacity induced by a weak magnetic transition, etc[3]. In recent years, the hunting for RTS even hotter than any time in the history of condensed matter physics. Some cases turn out to be scandals, but the faith of RTS persists. In below, we address some possible roadmaps for RTS.

**Metal hydrides under extreme high pressure**. The metallic hydrogen is one of the predicted RTS over 80 years ago. According to the Bardeen-Cooper-Schrieffer (BCS) theory, the electron pairs form by exchanging energies from phonon excitations and condense coherently. The hydrogen, as the lightest element in the universe, can form a metallic state under extreme high pressure, to support the strongest chemical bonding and high density of states on the Fermi surface. Unfortunately, the synthesis pressure of metallic hydrogen is close to 500 GPa, almost the upper limited hydrostatic pressure of diamond. Several reports on metallic hydrogen were published, but questions are more than evidences. Scientists turn to search superconductivity in hydrides, which may not require such pressure limit. This idea was first realized in $H_3S$ in 2015, then enormous binary hydrides are predicted and discovered, some of them show a possible superconducting transition above 200 K. Since hydrogen can easily bond with most of the elements in the periodic table in many forms, the hydrides give an endless gold mine for the hunting of RTS. Now, with data-driven strategy, people are searching for RTS in multinary hydrides even at ambient pressure [4].

**Unconventional superconductivity with pairing glues other than phonons**. In 1960s, G. M. Eliashberg, W. L. McMillan and P. W. Anderson predicted that the phonon-mediated Cooper pairs cannot form superconductivity above 40 K at ambient pressure, which is called as the McMillan limit. However, the discovery of high-$T_c$ superconductivity in cuprates breaks this limit in 1980s, so do the iron-based superconductors in 2008. Therefore, the cuprates and iron-based superconductors are categorized as unconventional superconductors, along with the heavy fermion superconductors, some Cr-based and Mn-based superconductors, and the newly discovered nickelates (Fig.1) [5]. Although the microscopic mechanism of unconventional superconductivity has not been well understood yet, more and more evidences suggest that the pairing glue of Cooper pairs is not pure phonon excitations. Some other candidates, such as magnetic fluctuations, polarons and plasmons are proposed, manifested by various scaling laws. With help from these new paring glues together with phonons, higher $T_c$ superconductivity may be discovered in future.

**Two-dimensional interface superconductivity**. Two-dimensional (2D) materials usually exhibit strong quantum fluctuations, which could play important roles in interface superconductivity. The classical example is $LaAlO_3$/$SrTiO_3$, while both of them are insulators, but the interface shows a superconductivity below $T_c$ =0.2 K. The bulk FeSe shows a superconductivity below $T_c$ =9 K, surprisingly $T_c$ in the interface between the single layer FeSe and $SrTiO_3$ (or $BaTiO_3$, $LaFeO_3$) is raised up to 65 K even 80 K. By choosing the right geometry in $(Bi,Sb)_2Te_3$/FeTe and EuO/$KTaO_3$ system, interface superconductivity emerges, too[2]. Following this idea, RTS is possible in combinations of interfaces.

**Tunable superconductivity by charge carriers**. The high-$T_c$ superconductivity can be tuned by chemical dopings, which probably introduce additional charge carriers or local crystalline distortion. This becomes a typical method to tune the conductivity in many 2D materials such as the twisted bilayer graphene. After further liquid or solid ionic gating, the existing superconductivity in low dimensional materials may be improved above 300 K.

**Nano-superconductors designed by atomic engineering**. Inspiring by the nano materials, we can design nano-superconductors by atomic engineering in many ways: different sandwich structures of 2D materials, nanotubes in 2D sheets, atomic clusters on the surfaces, 1D nanotubes with multiply walls, etc. This idea was also proposed in organic systems, theorists believed the complex nano-engineering may generate superconductivity even above 1000 K [2].

**Unexpected superconductors predicted by Artificial Intelligence (AI)**. Although there is no theoretical limitation on $T_c$ of unconventional superconductors, we do not have any theory to predict the RTS, either. This gives chances for AI hunting for RTS. By learning from the abundant data of existing superconductors and calculating the physical properties of unknow materials, AI could find unexpected matches without any theory, including the RTS [2].

**Superconductivity in highly disordered materials**. So far, superconductivity mostly emerges in crystalline solids. However, there are also some reports on the superconducting quasicrystals, high entropy alloys, even non-crystalline materials. In these materials, the long-range crystalline symmetry is broken, but Cooper pairs may survive in short range. Again, we do not have theories to describe the superconductivity in such disordered systems, and the possibility of RTS exists.

**Superconductivity in organic-inorganic hybrid materials**. If we extend our scope to organic-inorganic hybrid materials, there will be more candidates for RTS. Taking the organic ion intercalated FeSe superconductors $(CTA)_x FeSe$ and $(TBA)_x FeSe$ for examples, their $T_c$ reaches about 43 K, much higher than the bulk FeSe [2].

**Superconductors from outer space**. It is argued that metallic hydrogen may exist in the inner space of Jupiter due to the high pressure over 400 GPa. As we known, the solar system forms at billions years ago under extremely hot environment, high pressure and strong radiations. In the early ages, asteroid may contain superconducting compounds which cannot be artificially synthesized [2]. Superconductivity in meteorite was confirmed several years ago, but the $T_c$ is very low.

**Superconductivity with charge condensations other than electrons**. Instead, some astronomers proposed radical ideas about the neutron star, where its magnetic pulses probably originate from the flux excitations of superconducting state of protons. Such proton superconductors may host a $T_c$ up to 10 billion K, but it is impossible to testify this picture [2]. However, in some materials, the charge carriers may not electrons but other particles or quasiparticles, the high velocity and near massless nature of them may induce superconductivity near 300 K.

The RTS is believed as a jewel in the crown of condensed matter physics, but we should be very cautious and keep claim for the hunting of RTS. Even the RTS is discovered, it may be useless in high power electricity transport or high field applications for the possible low $J_c$ and $H_{c2}$. No matter what the future of RTS, the pursuing for RTS requires the spirit of exploration, endeavor and curiosity in science, it will certainly change our world and knowledge forever.

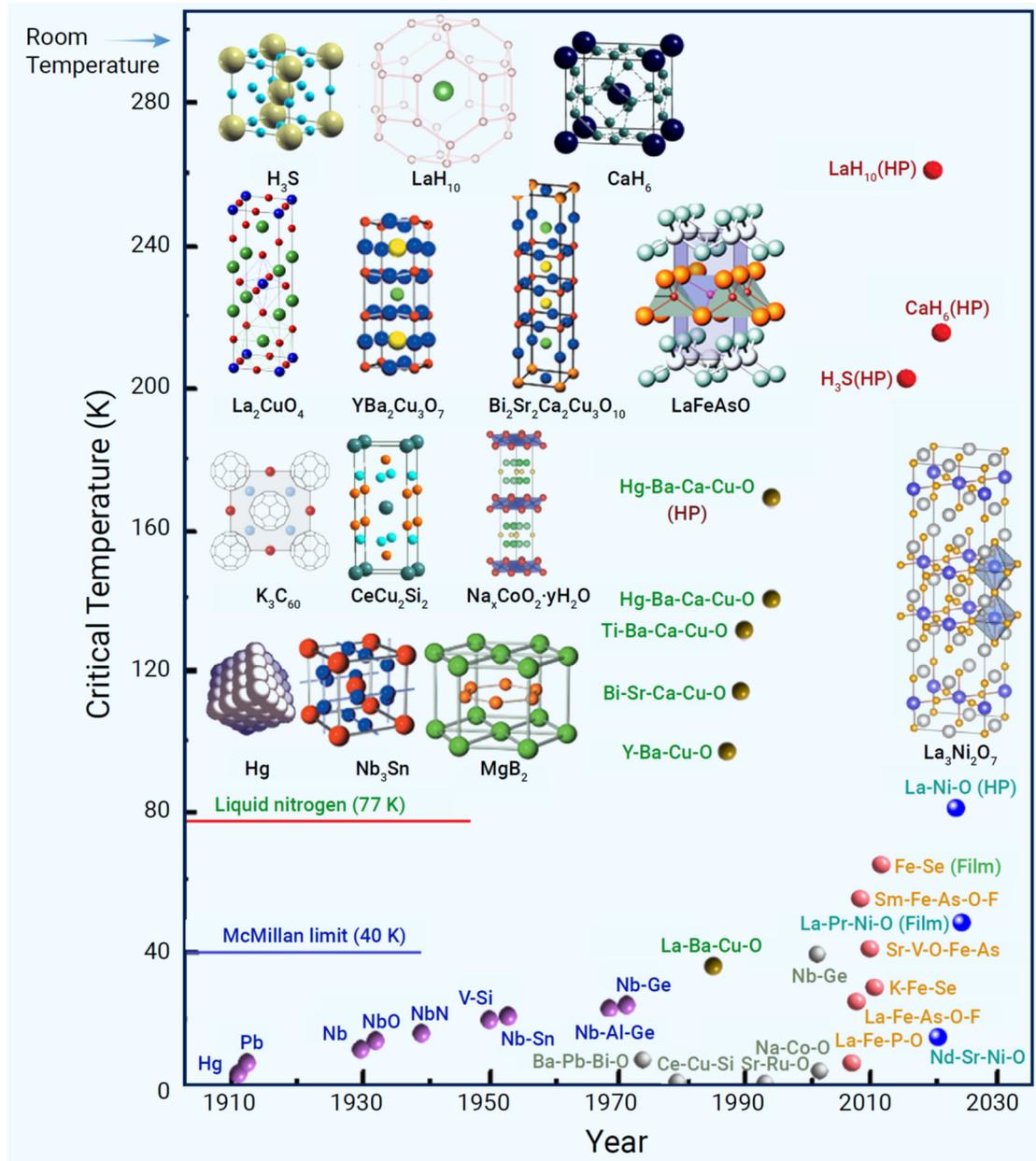

Figure 1. The discovery of typical superconductors and their critical temperature.

superconductivity in clathrate hydrides at high pressure. The Innovation **3**(2), 100226.
5. Qin,Q., Wang, J. and Yang,Y. -f., (2024). Frustrated superconductivity and intrinsic reduction of Tc in trilayer nickelate. The Innovation Materials **2**, 100102.

**ACKNOWLEDGMENTS**

This work is supported by the National Key Research and Development Program of China (Grant Nos. 2023YFA1406100 and 2018YFA0704200), the National Natural Science Foundation of China (Grant Nos. 11822411, 11874057 and 11961160699), the Strategic Priority Research Program (B) of the CAS (Grant No. XDB25000000) and the Youth Innovation Promotion Association of CAS (Grant No. Y202001).

**DECLARATION OF INTERESTS**

The author declares no competing interests.